\documentclass[aps,pra,reprint,superscriptaddress,amsmath]{revtex4-1}
\usepackage{graphicx}
\usepackage[utf8]{inputenc}
\usepackage{hyperref}
\usepackage{color}
\newcommand{\vbf}[1]{{\mathbf #1 }}

\newcommand{\ket}[1]{\left| #1 \right\rangle }

\newcommand{\braket}[2]{\left\langle #1 \right| \left. #2 \right\rangle }

\newcommand{\bs}[1]{\mathcal{#1}}

\begin{document}

\preprint{}

\title{Modeling the adiabatic creation of ultracold, polar $\mathrm{^{23}Na^{40}K}$ molecules}

\author{Frauke Seeßelberg}
\email[]{frauke.seesselberg@mpq.mpg.de}
\affiliation{Max-Planck-Institut für Quantenoptik, 85748 Garching, Germany}
\author{Nikolaus Buchheim}
\affiliation{Max-Planck-Institut für Quantenoptik, 85748 Garching, Germany}
\author{Zhen-Kai Lu}
\affiliation{Max-Planck-Institut für Quantenoptik, 85748 Garching, Germany}
\author{Tobias Schneider}
\affiliation{Max-Planck-Institut für Quantenoptik, 85748 Garching, Germany}
\author{Xin-Yu Luo}
\affiliation{Max-Planck-Institut für Quantenoptik, 85748 Garching, Germany}
\author{Eberhard Tiemann}\affiliation{Institut für Quantenoptik, Leibniz Universität Hannover, 30167 Hannover, Germany}
\author{Immanuel Bloch} 
\affiliation{Max-Planck-Institut für Quantenoptik, 85748 Garching, Germany}
\affiliation{Fakultät für Physik, Ludwig-Maximilians-Universität München, 80799 München, Germany}
\author{Christoph Gohle}
\affiliation{Max-Planck-Institut für Quantenoptik, 85748 Garching, Germany}

\date{\today}

\begin{abstract}
In this work we model and realize stimulated Raman adiabatic passage (STIRAP) in the diatomic $\mathrm{^{23}Na^{40}K}$ molecule from weakly bound Feshbach molecules to the rovibronic ground state via the $\ket{v_d=5,J=\Omega=1}$ excited state in the $d^3\Pi$ electronic potential.
We demonstrate how to set up a quantitative model for polar molecule production by taking into account the rich internal structure of the molecules and the coupling laser phase noise.
We find excellent agreement between the model predictions and the experiment, demonstrating the applicability of the model in the search of an ideal STIRAP transfer path. 
In total we produce 5000 fermionic groundstate molecules.
The typical phase-space density of the sample is 0.03 and induced dipole moments of up to 0.54 Debye could be observed.
\end{abstract}

% insert suggested PACS numbers in braces on next line
% insert suggested keywords - APS authors don't need to do this
%\keywords{}

\maketitle

Dipolar quantum gases allow for the realization of intriguing new quantum many-body systems and associated phenomena due to their anisotropic and long-range interactions.
Among these are the roton driven fluid to crystalline quantum phase transition \cite{Chomaz2017}, dipolar droplet formation \cite{Kadau2015,Ferrier_Barbut2016}, insulators with fractional filling and supersolid phases of dipoles in optical lattices \cite{Capogrosso-Sansone:2009eu} to name only a few. 
Ultracold polar molecules promise particularly large dipolar interactions due to their large dipole moments. 

The standard procedure for creating molecules at high phase-space density starts with a mixture of two atomic species close to quantum degeneracy. 
The two species are then initially adiabatically associated into a weakly bound Feshbach molecular state $\ket{FB}$ \cite{Ferlaino2008}. 
From there they can be transferred into the final, electronic, vibrational and rotational (rovibronic) ground state using stimulated Raman adiabatic passage (STIRAP) \cite{Bergmann1998,Vitanov2017}. 
This last step involves coupling the initial and final state to a common intermediate, electronically excited molecular state.
Both $\ket{FB}$ and the intermediate state need to be chosen with care in order to allow for a high efficiency in the transfer and thus to preserve the phase space density of the ultracold mixture.
This approach has been applied successfully to dipolar KRb \cite{Ni2008}, RbCs \cite{Takekoshi2014,Molony2014}, NaK \cite{Park2015} and NaRb \cite{Guo2016} molecules.

Here we demonstrate the transfer $^{23}$Na$^{40}$K Feshbach molecules, created close to the $m_F=-7/2$ Feshbach resonance at 88~G \cite{Park2012}, via the $\ket{v_d=5,J=\Omega=1}$ state, associated with the $d^3\Pi$ potential, to the rovibronic ground state.
Here $v_d$ refers to the vibrational quantum number of a level associated with the $d^3\Pi$ potential, $J$ is the total angular momentum of the molecule excluding nuclear spins and $\Omega$ is the projection of $J$ onto the internuclear axis. 
This intermediate state, which has unresolved hyperfine (HF) structure, provides an alternative route to the ground state compared to the STIRAP scheme employed in \citep{Park2015}. 
We develop a Hamiltonian model to describe the adiabatic transfer in all required details to achieve a quantitative description. 
In addition to the molecular structure analysis done for different bialkali systems \cite{Schulze2013, Park2015a} we include the complex light coupling into the analysis.
This results in a multi-level, cross-coupled model that is intimately related to the work of the Bergmann group on STIRAP in multilevel systems \cite{Shore1995} but is specific to the alkali-alkali molecule formation. 
We investigate how to maximize the STIRAP transfer efficiency for a given intermediate state manifold by optimizing pulse durations and one-photon detuning. 
We find excellent agreement between simulation and experiment. 
Finally, we demonstrate ground state molecule creation with a large electric dipole moment of up to 0.54 Debye.

\section{Molecular level structure and Hamiltonian model}
In our model we use a STIRAP coupling field $\vbf{E}(t)$ of the form (see e.g. \cite{Yatsenko2014})
\begin{eqnarray}
\vbf{E}(t) &=& \vbf{E}_{P}(t) \sin(\omega_P t + \phi_P(t))\nonumber\\
&+&\vbf{E}_{S}(t) \sin(\omega_S t + \phi_S(t))\label{eq:pulse}\\
\vbf{E}_{P}(t) &=& \vbf{E}_{0,P} \sin  \left(\frac{\pi}{2} \frac{t}{\tau}\right),\quad\vbf{E}_{S}(t) = \vbf{E}_{0,S} \cos  \left(\frac{\pi}{2} \frac{t}{\tau}\right)\nonumber
\end{eqnarray}
where $\vbf{E}_{0,x}$ denotes the amplitude vector, $\phi_x(t)$ a time dependent phase (noise) term and $\omega_x$ the carrier frequency, the index $x$ distinguishing between either pump ($P$) or Stokes ($S$) field. 
$\tau$ is the STIRAP pulse duration.
We work in the rotating frame of these laser fields and employ the rotating wave approximation (RWA).  
Coupling matrix elements between two states $\{i,j\}$ take the form $\Omega_{i,j} = \vbf{E}_{x}(t)\vbf{d}_{i,j} e^{\pm i\phi_{x}(t)}$. Here $\vbf{d}_{i,j}$ is the corresponding transition dipole moment.
The sign in the exponent and $x$ are determined according to the RWA.

\begin{figure*}
\includegraphics{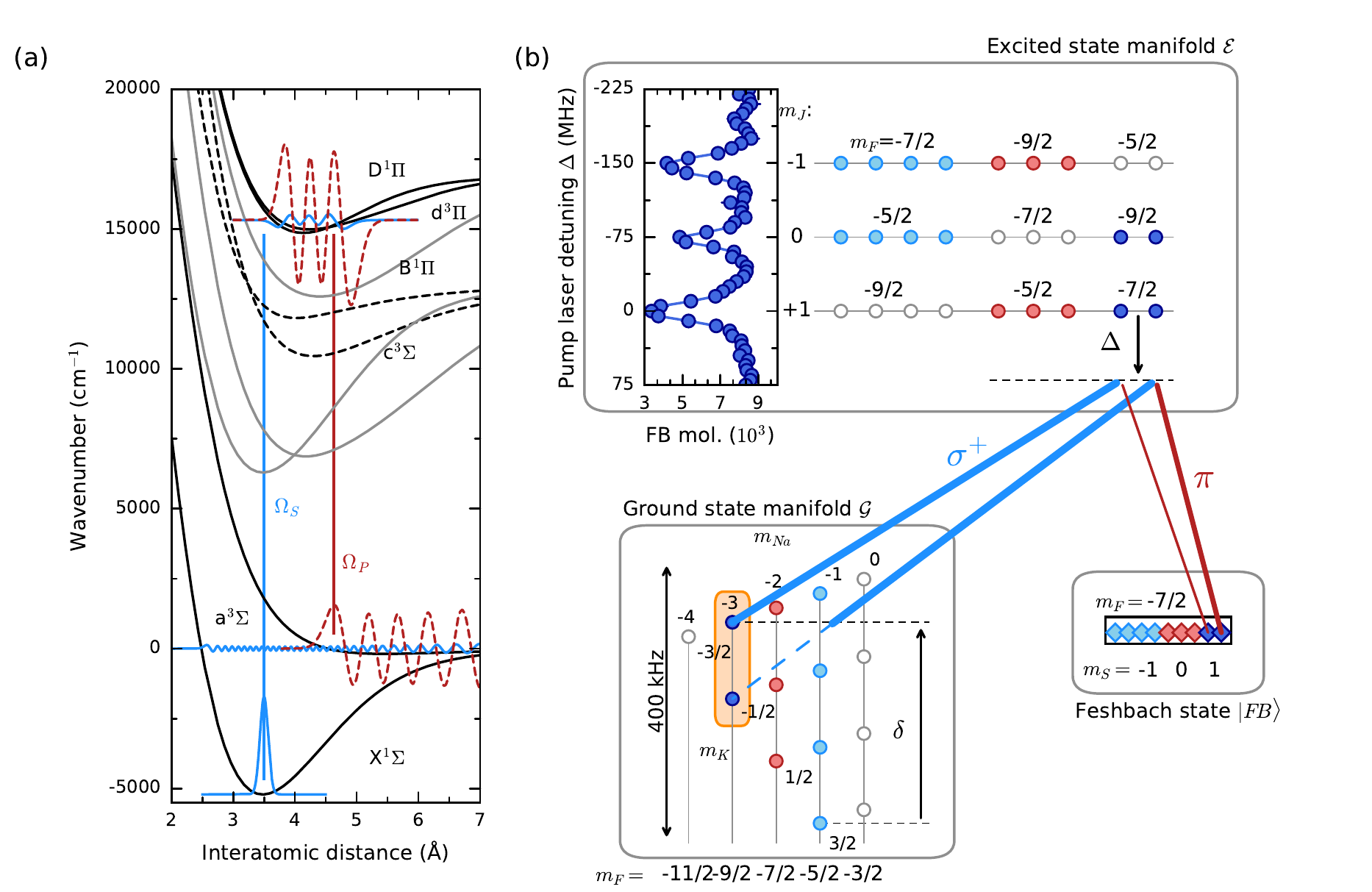}
\caption{\label{fig:schematic}Molecular structure of $^{23}$Na$^{40}$K. 
(a) 
Potential energy curves according to \cite{Aymar2007}: 
The ground state potentials X$^1\Sigma$ and a$^3\Sigma$ and the excited state potentials D$^1\Pi$ and d$^3\Pi$, which are relevant for the present work, are highlighted by black solid lines. 
The B$^1\Pi$/c$^3\Sigma$ system used in \citep{Park2015} is indicated with dashed lines. 
The vertical lines symbolize the pump- (P) and Stokes (S) lasers used to populate the rovibronic ground state by STIRAP, with the respective Rabi frequencies denoted as $\Omega_P$ and $\Omega_S$. 
Exemplary we show the singlet (solid) and triplet (dashed) component of one state in $\bs{E}$ and $\bs{G}$ and one spin projection for $\ket{FB}$ (scaled up by a factor of 100).  
(b) 
Schematic of the molecular structure of the levels involved in STIRAP. 
The experimental data on the upper left shows the spectrum of the excited state $v_d=5$, $J=\Omega=1$ at 85.5 G recorded with 45$^{\circ}$ polarization and starting with the $m_F = -7/2$ Feshbach molecules that can be created at this field (lines are a guide to the eye). 
Three Zeeman $m_J$-components are clearly visible, but no hyperfine structure is resolved. 
The individual hyperfine states of the excited and the ground state are indicated schematically (circles) as well as the hyperfine components of the Feshbach state (diamonds). 
Symbols with the same total nuclear spin quantum number $m_I=m_{Na}+m_K$ (-5/2,-7/2,-9/2) have the same color (light blue, red, dark blue); white symbols refer to states that are not populated.
Exemplary, shown by arrows, is the case of a $\pi$-polarized pump beam and a $\sigma^{+}$-polarized Stokes beam. 
In this case, only the two $m_F$=-7/2 components of $m_J=1$ contribute to STIRAP. 
The strengths of the pump and Stokes transitions are different, as indicated by the thickness of the arrows.  
The one-photon detuning $\Delta$ and the two-photon detuning $\delta$ are also indicated. 
Note that the energy axis for the excited state (spectrum and schematic) is inverted for clarity.}
\end{figure*}

The initial state for our STIRAP process, which is illustrated in Fig. \ref{fig:schematic}, is the weakly bound molecular state associated with the Feshbach resonance at $88.9$~G that we create at a magnetic field $B=B_F=85.5$~G. 
We denote its state vector as $\ket{FB}$ and define its energy as $E_{FB}=0$.
We ignore all other states in the vicinity of $\ket{FB}$, as they are energetically far detuned from $\ket{FB}$.

Within the manifold of the rovibronic ground states, $\bs{G}$, the nuclear spin is the only degree of freedom and the only contribution to the magnetic moment of bialkali molecules.
Therefore the Hamiltonian contains only nuclear Zeeman and nuclear spin-spin interaction terms. 
For $^{23}$Na$^{40}$K with $I_{Na}=3/2$ and $I_K=4$, the nuclear spin basis contains $4\times9=36$ states. 
At $B_F$, the Hamiltonian of $\bs{G}$ can be approximated in the Paschen-Back limit, which is justified because of small HF interactions, as
\begin{eqnarray*}
\hat{H}_\bs{G}/\hbar  &=& [(\hat I_{z,Na}-3/2) \mu_{Na} + (\hat I_{z,K}+4) \mu_K]B + \\
&+&c_4 \hat I_{z,Na} \hat I_{z,K} - \delta,
\end{eqnarray*}
where $\mu_{Na/K}$ denote the magnetic moments of the sodium and potassium nuclei, $\hat I_{z,Na/K}$ are the projections of the nuclear spin operators of the respective nucleus onto the magnetic field axis with eigenvalues $m_{Na/K}$ and
$c_4 \approx 2 \pi\times 0.4$~kHz is the scalar spin-spin interaction constant \citep{Park2015,Park2016}. 
The two-photon detuning $\delta$ of the coupling lasers is defined relative to the HF ground state at field $B_F$ with energy $E_\bs{G}$
% ($\ket{m_{Na},m_{K}}=\ket{3/2,-4}$, energy $E_G$)
, i.e. $\delta=E_\bs{G}/\hbar-(\omega_S-\omega_P)$. 
Other molecular states are detuned by at least twice the rotational constant in the ground state, $2B\approx 5.6$~GHz, and can safely be ignored. 

For the NaK system the $\ket{v_d=5,J=\Omega=1}$ is a suitable intermediate state manifold $\bs{E}$ for STIRAP \cite{Nadia}. 
It has significant spin-orbit coupling that results in a 2\%  admixture of the $D^1\Pi, v_D=6$ state and suitable transition dipole moments to $\bs{G}$ and $\ket{FB}$ with a magnitude on the order of 0.01 D.
In particular the pump transition dipole moment is about one order of magnitude larger than for the previously used $\ket{v_c=35,J=1}$ state \cite{Aymar2007, Park2015a}.
The pump transition matrix element limits the maximal coupling in STIRAP since it is smaller than the Stokes transition matrix element. 
For our intermediate state however the electronic spin projection $\Sigma=0$ vanishes resulting in the absence of a Fermi contact HF interaction and since the orbital interaction in Na is small \cite{Townes1955} the HF structure in $\bs{E}$ can not be resolved spectroscopically in the present study. 
Therefore we approximate the Hamiltonian of $\bs{E}$ by a pure Zeeman term and an imaginary damping term, that models decay to other molecular states
\begin{equation*}
\hat{H}_\bs{E}/\hbar = (\hat{J}_z-1) g \mu_B B - \frac{i \gamma}{2} - \Delta,
\end{equation*}
$\hat{J}_z$ denotes the angular momentum operator along the magnetic field axis with eigenvalues $m_J$ and $\mu_B$ is the Bohr magneton. 
For Hund's case (a) $g=\Omega(\Lambda+g_e\Sigma)/(J(J+1))$ \cite{Townes1955}, where $g_e$ denotes the g-factor of the electron, so that $g=1/2$ for this state.  
The excited states decay with a rate $\gamma$ and $\Delta = \omega_P-E_\bs{E}/\hbar$ is the detuning of the pump laser from the transition from $\ket{FB}$ to the upper Zeeman component $\ket{\bs{E},m_J=1}$ with energy $E_\bs{E}$. 
The total number of states in this manifold with $J=1$ is $36\times3 = 108$. 
No further molecular levels have to be considered, since even the nearest one $\ket{v_d=5,J=2, \Omega=1}$ is already $7.2$~GHz away.
A damped Hamiltonian evolution is a good approximation to the full dynamics since spontaneous decay from this intermediate state ends almost exclusively in uncoupled states. 

To express the described Hamiltonian as a matrix, we employ a nuclear spin decoupled molecular basis $\{\ket{FB},\ket{n, J, m_J, m_{Na}, m_{K}}\}$, where $n\in\{\bs{E,} \bs{G}\}$. 
Note that the physical meaning of $J$, the total angular momentum without nuclear spins, depends on $n$: 
For the Feshbach molecule $J$ is equal to the total electronic spin $S\in\{0,1\}$, for the excited state $J=1$ and in the ground state $J=0$.  
In this basis the Hamiltonian is diagonal. 
It is therefore convenient to expand the Feshbach state $\ket{FB}$ in terms of the same spin basis
\begin{equation*}
\ket{FB}=\sum_{J,m_J,m_{Na},m_K} \ket{J m_J m_{Na} m_{K}}\ket{\Psi_{FB, J m_J m_{Na} m_{K}}},
\end{equation*}
where $\ket{\Psi_{FB, J m_J m_{Na} m_{K}}}$ are the radial parts of the projection of $\ket{FB}$ on to the respective spin states.
The excited states in $\bs{E}$ are given by
\begin{equation*}
\ket{\bs{E},J m_J m_{Na} m_{K}} = \ket{\bs{E}, J m_J} \ket{m_{Na}}\ket{m_K}(\ket{\Psi_{\bs{E},0}}+\ket{\Psi_{\bs{E},1}}),
\end{equation*}
where $\ket{\Psi_{\bs{E},s}},  s\in\{0,1\}$ are the electronic spin singlet and triplet components of the radial part of $\bs{E}$. 
In contrast to $\ket{FB}$, they essentially do not depend on the nuclear spin states because of negligible HF interaction compared to spin-orbit interaction. 
Finally the HF states in $\bs{G}$ are given by
\begin{equation*}
\ket{G, m_{Na} m_K} = \ket{G, J=0, m_J=0} \ket{m_{Na}} \ket{m_K}  \ket{\psi_\bs{G}},
\end{equation*}
with $\ket{\psi_\bs{G}}$ being the radial part of the ground state wavefunction.

Next we determine the coupling matrix elements relevant for STIRAP.
For the pump transition $\ket{FB}\rightarrow\bs{E}$ it is proportional to
\begin{align}
\label{eq:pumpmatrix}
&\braket{FB|\mathbf{E}\cdot \hat{\mathbf{d}}}{\bs{E},J m_J m_{Na} m_{K}} \\
&\propto E_P \sum_{J' m'_J m'_{Na} m'_{K} q} \alpha_q (2J+1)^{-1/2} \nonumber\\
&\times \braket{J' m'_J 1 q}{J m_J} \nonumber\\
&\times \braket{\Psi_{FB, J' m'_J m'_{Na} m'_{K}}}{\Psi_{\bs{E}, J}} \nonumber\\ 
&\times \braket{m'_{Na}}{m_{Na}} \braket{m'_K}{m_K} \nonumber 
\end{align}
where $q$ labels the polarization (0 corresponds to $\pi$-polarization and $\pm1$ to $\sigma^{+}$/$\sigma^{-}$) and $\alpha_q$ is the polarization vector of $\vbf{E}_P(t)$. 
In Eq. (\ref{eq:pumpmatrix}) the first factor is the conventional Clebsch-Gordan coefficient and represents the part of the Hönl-London factor which depends on the laboratory fixed quantum numbers, the second factor is the radial function overlap integral, the square of which is the Franck-Condon (FC) factor, and the last ones are matrix elements in the nuclear spin space yielding zero or one.  
We apply the Franck-Condon principle assuming that the electronic transition moment is constant over the needed internuclear separation.
To obtain a sufficiently accurate $\ket{FB}$ wavefunction, a coupled channel calculation \cite{Temelkov2015} for this molecular state was performed.
For the chosen intermediate state, the FC factors in the above expression originate mainly from the inner turning point of the triplet part of the Feshbach wavefunction, see Fig. \ref{fig:schematic}(a). 
Since the singlet part with $J'=0$ is rapidly oscillating, its FC factors are very small and thus all singlet terms will be neglected in the coupling between $\ket{FB}$ and $\bs{E}$. 
\begin{table}
\caption{\label{tab:Francks} Overlap integrals for the pump transition for different spin components of the Feshbach molecule for $m_F = -7/2$ and $S=1$. The sum of the squared values is normalized to 1.}
\begin{ruledtabular}
\begin{tabular}{c c c c c}
$m_J$ & $m_{Na}$ & $m_K$ & overlap integral \\ \hline
-1 & -3/2 & -1 & -0.095 \\
-1 & -1/2 & -2 & 0.209 \\
-1 & 1/2 & -3 & 0.148 \\
-1 & 3/2 & -4 & 0.114 \\ \hline
0 & -3/2 & -2 & -0.100 \\
0 & -1/2 & -3 & -0.223 \\
0 & 1/2 & -4 & -0.708 \\ \hline
1 & -3/2 & -3 & 0.362 \\
1 & -1/2 & -4 & 0.470 \\
\end{tabular}
\end{ruledtabular}
\end{table}
Overlap integrals for our specific $\ket{FB}$ and $\bs{E}$ are given in table \ref{tab:Francks}.

For the Stokes transition $\bs{E} \rightarrow \bs{G}$ the transition matrix element is
\begin{align}
&\braket{\bs{E},J m_J m_{Na} m_{K}|\mathbf{E}\cdot \hat{\mathbf{d}}}{\bs{G}, m'_{Na} m'_K} \\
& \propto E_S(t) \sum_q \beta_q \braket{J m_J 1 q}{0 0} \nonumber\\
&\times \braket{m_{Na}}{m'_{Na}} \braket{m_K}{m'_K}\braket{\psi_{E,1}}{\psi_\bs{G}},\nonumber
\end{align}
where $\beta_q$ is the polarization vector of the Stokes field $\vbf{E_S(t)}$. 
Since the nuclear spins factorize everywhere, we can reduce our nuclear basis to only the 9 components present in the $\ket{FB}$ state (Tab. \ref{tab:Francks}). 
While $\bs{G}$ is the angular momentum singlet ($J=0$), $\bs{E}$ is a triplet ($J=1$) and therefore the maximal size of the basis is $(1+3)\times9+1=37$ states. The entire Hamiltonian matrix is represented graphically in Fig. \ref{fig:schematic}(b) with the dominant coupling terms.

Tab. \ref{tab:Francks} shows that the largest coupling matrix elements are those involving the $(J,m_J,m_{Na},m_{K}) = \{(1,0,1/2,-4), (1,1,-1/2,-4), (1,1,-3/2,-3)\}$ spin projections of $\ket{FB}$.
For resonant driving ($\Delta=0$) the dynamics will be dominated by couplings to $m_J=1$ states in $\bs{E}$. 
With $\pi$ polarization on the pump field (scenario A) those are coming from the two $m_J=1$ projections of $\ket{FB}$ which are indicated with dark blue diamonds in Fig. \ref{fig:schematic}(b). 
Similarly using $\sigma^+$ on the pump field (scenario B), the nuclear spin projection $m_{Na} = 1/2, m_K = -4$ plays the largest role. In both cases the Stokes field has to have $\sigma^+$ polarization. 

\section{Experimental results}
Our experimental setup produces ultracold mixtures of bosonic $^{23}$Na and fermionic $^{40}$K. 
We prepare {\raise.17ex\hbox{$\scriptstyle\mathtt{\sim}$}} $1.3 \times 10^5$ atoms of each species in a crossed, far-detuned optical dipole trap at a temperature of 0.7 $\mu$K, the phase space density of the sample being about 0.5. 
Sodium is prepared in the $\ket{F,m_F}=\ket{1,1}$ state and potassium in the $\ket{9/2,-7/2}$ state before we ramp up the magnetic field to 85.5 G, close to an interspecies Feshbach resonance located at 88 G in the $m_{F,Na}=1, m_{F,K}=-9/2$ collision channel \cite{Park2012}. 
We use a radio-frequency sweep to flip the potassium atoms into the $m_F=-7/2$ molecular bound state associated with the Feshbach resonance. 
The efficiency of this process is roughly 10 \% and we typically create {\raise.17ex\hbox{$\scriptstyle\mathtt{\sim}$}}$1.1\times 10^4$ Feshbach molecules with a binding energy of 80 kHz.  
For the STIRAP, lasers with wavelengths of 652~nm (pump) and 487~nm (Stokes) are required, for which we use a diode- and a dye-laser, respectively. 
Both lasers are phase-locked to master diode-lasers, which in turn are locked to the same ultrastable Fabry-Perot reference cavity and have sub-kHz linewidths. 
The beams propagate perpendicular to the magnetic field axis so that we can realize parallel ($\pi$) or perpendicular ($\perp\equiv (\sigma^++\sigma^-)/\sqrt{2}$) polarization.

\begin{figure}
\includegraphics{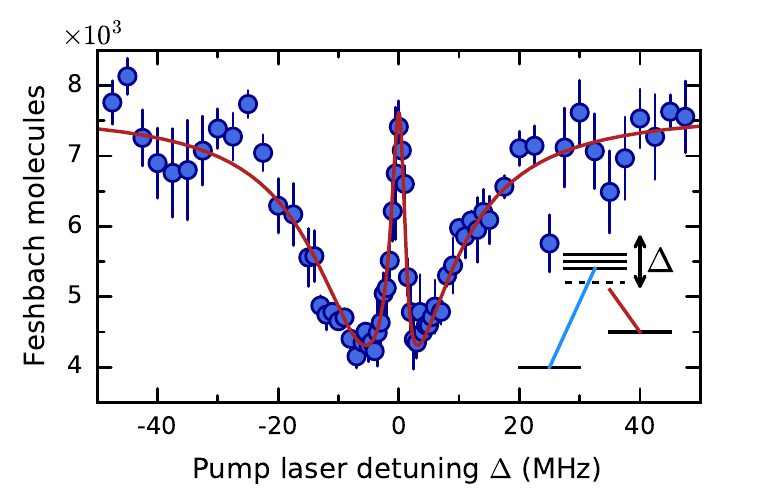} 
\caption{\label{fig:EIT} 
EIT spectrum as measured in the experiment (circles) by scanning the pump laser detuning $\Delta$ while keeping the Stokes laser resonant with $m_J=1$ component of the excited state, $\delta=\Delta$. 
Error bars denote standard deviations of several experimental runs. 
The line is a fit using Eq. (\ref{eq:eit}). 
From this fit we extract a peak pump Rabi frequency $\Omega_P = 2\pi \times 3.9$~MHz and a peak Stokes Rabi frequency $\Omega_S = 2\pi \times 8.4$~MHz. 
See text for details.
}
\end{figure}

We image Feshbach molecules directly using absorption imaging. 
The absorption cross section remains essentially unchanged compared to atoms.
We calibrate our pump and Stokes field strengths by recording a spectrum on the $\ket{FB}\rightarrow\ket{\bs{E},J=1, m_J=1}$ transition with weak pump and resonantly tuned Stokes fields (Fig. \ref{fig:EIT}). 
The profile is fit using a three level model for electromagnetically induced transparency (EIT) \cite{Fleischhauer2005a} 
\begin{equation}
\label{eq:eit}
N \propto \exp\left(-t \Omega_P^2 \frac{4\gamma\delta^2}{|\Omega_S^2 + 2i\delta (\gamma + 2 i \Delta)|^2}\right),
\end{equation} 
where $t$ is the EIT pulse duration and $\gamma$ the excited state line width. 
 
From the fit, we obtain a linewdith of $\gamma = 2\pi\times20$~MHz and Rabi frequencies $\Omega_S = 2\pi\times8.4$~MHz and $\Omega_P = 2\pi\times2.6$~MHz with a total power of $10$~mW ($100$~mW) for the Stokes (pump) beams and a spot size of $w\approx18$~$\mu$m at the atoms. 
Since the Stokes matrix elements do not depend on the nuclear spin and the transparency peak is much wider than the ground state energy spread (see Fig. \ref{fig:schematic}(b)), we can directly use $\Omega_S$ as the peak Rabi frequency for those matrix elements. 
To account for all excitation paths on the pump transition, we adjust $E_P$ in (\ref{eq:pumpmatrix}) such that $\sum_{m_{Na},m_{K}}\left|\braket{FB|\mathbf{E}\cdot \hat{\mathbf{d}}}{\bs{E},J=1, m_J=1, m_{Na} m_{K}}\right|^2=\Omega_P^2$.

To perform STIRAP, we use pulses with the power envelope of Eq. (\ref{eq:pulse}) with smooth turn on and off.
We reverse the STIRAP pulse sequence after a hold time of 90~$\mu$s.
During this time we remove remaining potassium atoms from the trap using a resonant light field
to obtain a background free STIRAP signal. 
For the following experiments we determine the ground state molecule signal as the ratio of the Feshbach molecules after and before this procedure, denoted as the round trip fraction $\eta^2$. 
The STIRAP efficiency is thus $\eta$, assuming that both STIRAP processes are equally efficient.
\begin{figure}
\includegraphics{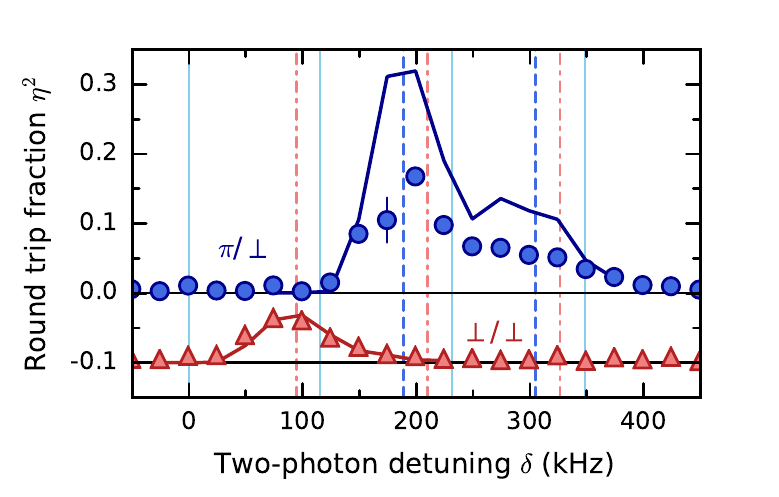}
\caption{\label{fig:groundstate} 
Hyperfine spectra of the rovibronic groundstate. 
They were recorded with different polarizations at $\Delta=100$~MHz and using 70~$\mu$s STIRAP pulse duration. 
Circles (triangles) denote the data recorded with $\pi$($\perp$)-pump polarization, the Stokes beam being always $\perp$-polarized.
Error bars denote standard error of mean of several experimental runs and are mostly smaller than the symbol size. 
Vertical lines indicate the positions of the lowest hyperfine levels ($m_F=-5/2$ solid, $-7/2$ dashed, $-9/2$ dash-dotted, same color convention as in Fig. \ref{fig:schematic}(b). 
Simulation results for the two polarization scenarios are indicated by solid lines, for which the experimentally measured phase noise was included (see text for details).
The data and simulation result for the $\perp$/$\perp$ polarization configuration are offset by -0.1 for clarity. 
}
\end{figure}

By scanning $\delta$ we observe spectral structures that correspond to HF states in $\bs{G}$, see Fig. \ref{fig:groundstate}. 
Here $\Delta=100$~MHz and $\tau=70$~$\mu$s. 
Depending on the polarization of the STIRAP beams, different ground states can be populated.
First we work with a $\perp$ polarized Stokes field while the pump field has $\pi$ polarization, so almost scenario A. 
In this case, we observe the nuclear spin states in the $m_S=1$ subspace of $\ket{FB}$ (circles).
The largest STIRAP efficiency $\eta$ is then obtained for the $\ket{m_{Na},m_{K}}=\ket{-1/2,-4}$ hyperfine state at $\delta=200$~kHz, consistent with Tab. \ref{tab:Francks}.

\begin{figure}[!h]
\includegraphics{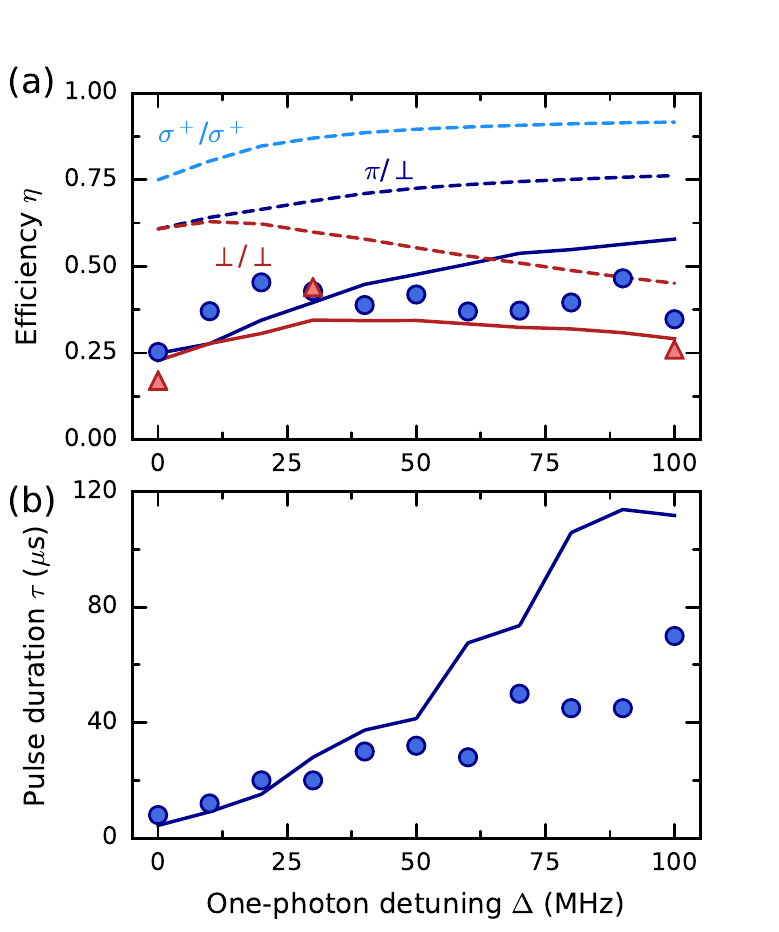} 
\caption{\label{fig:effi} 
STIRAP at different one-photon detunings.
(a) 
STIRAP efficiency as measured in the experiment (symbols), error bars denoting the standard error of the mean are smaller than the symbol size, model results without phase noise (dashed lines) and including phase noise (solid lines, averaged over 12 simulations) are also shown. 
Color and symbol shape encode polarization scenarios. 
Circles and dark blue lines (triangles and red lines) denote $\pi$($\perp$)-pump polarization, the Stokes beam is always $\perp$-polarized. 
Light blue refers to the experimentally not realized case of $\sigma^+$ polarization for both pump and Stokes laser. 
(b) 
Optimal STIRAP duration $\tau$ as determined for each $\Delta$ for the $\pi$/$\perp$ situation. 
Data (circles) with model prediction including phase noise (solid line).
}
\end{figure}
To optimize the STIRAP process we investigate the transfer efficiency $\eta$ to the $\ket{-1/2,-4}$ HF ground state for different one-photon detunings $\Delta$, optimizing $\tau$ for each value of $\Delta$ assuring two-photon resonance $\delta=0$. 
The result is shown in Fig. \ref{fig:effi}(a) (circles). 
We find that the efficiency is 25\% for one-photon resonant STIRAP, but rises up to {\raise.17ex\hbox{$\scriptstyle\mathtt{\sim}$}}50\% for detunings larger than 20 MHz and then saturates. 
Also shown is the result of the parameter free calculation (dashed dark blue line) for optimal pulse duration according to the model, neglecting noise. 
The behaviour can be qualitatively understood by realizing that scattering from unwanted components decreases as $1/\Delta^2$ while coupling only decreases as $1/\Delta$, which can be compensated with longer pulse durations.
Note, that the ideal model predicts a significantly larger efficiency than the one observed in the experiment.
However, when we include a realistic phase noise function $\phi_x(t)$ into the model, we can resolve this discrepancy: 
In order to do so, we apply a random $\phi_x(t)$ that reproduces the measured beat note radio-frequency spectrum between each STIRAP laser to their respective master laser. 
The phase noise power spectrum has a bandwidth of about $2.5$~MHz and a magnitude that yields an rms amplitude $\phi_{x,\mathrm{rms}} = 400$~mrad. 
This noise function is multiplied by a factor $\sqrt{2}$, assuming the phase noise of the master laser to the cavity lock is the same as the phase noise of the slave laser to the master lock. 
Including the laser phase noise spectra into the model calculation leads to the solid dark blue line that matches the data fairly well.
It can be seen that the influence of the phase noise on the molecule production is strongest close to resonance and becomes less prominent for larger $\Delta$. 
Fig. \ref{fig:effi}(b) shows the optimal STIRAP pulse durations $\tau$, both obtained from the experiment (circles) and the model including phase noise (solid line). 
Also in this case the model describes accurately what we observe.
Both the observed efficiency and the ideal STIRAP pulse duration agree very well for small $\Delta$.
At large $\Delta$ experimentally optimal pulses are shorter.
This indicates, that for larger $\Delta$ with a reduced effective two-photon coupling and longer pulses other noise sources may become important. 
This is also consistent with larger predicted efficiency at large $\Delta$.
We can also compare the HF spectra of Fig. \ref{fig:groundstate} with our model:
Using experimental parameters including phase noise the modeled spectra match. Only the amplitude for the $\pi/\perp$ case is systematically too large.

To further benchmark the accuracy of the model calculation, we study STIRAP in a second polarization scenario, where pump and Stokes beam are both $\perp$ polarized.
This is not quite scenario B as discussed before, as the $\sigma^-$-component can also couple to excited state components.
Still also in that case mainly the $\ket{1/2,-4}$ HF state is populated, see Fig. \ref{fig:groundstate} (triangles).
The corresponding efficiency measurements are indicated with triangles in Fig. \ref{fig:effi}(a).
And also in this case detuned STIRAP is favorable compared to resonant STIRAP.

Finally we simulate the ideal polarization scenario, scenario B, that could not yet be implemented experimentally due to geometrical constraints in the experimental apparatus, requiring $\sigma^+$/$\sigma^+$ polarized pump/Stokes beams.
This scenario also addresses the $\ket{1/2,-4}$ ground state and according to the simulation should yield the highest transfer efficiencies (light blue line in Fig. \ref{fig:groundstate}(a)) of all three polarization scenarios discussed. 
 
\begin{figure}
\includegraphics{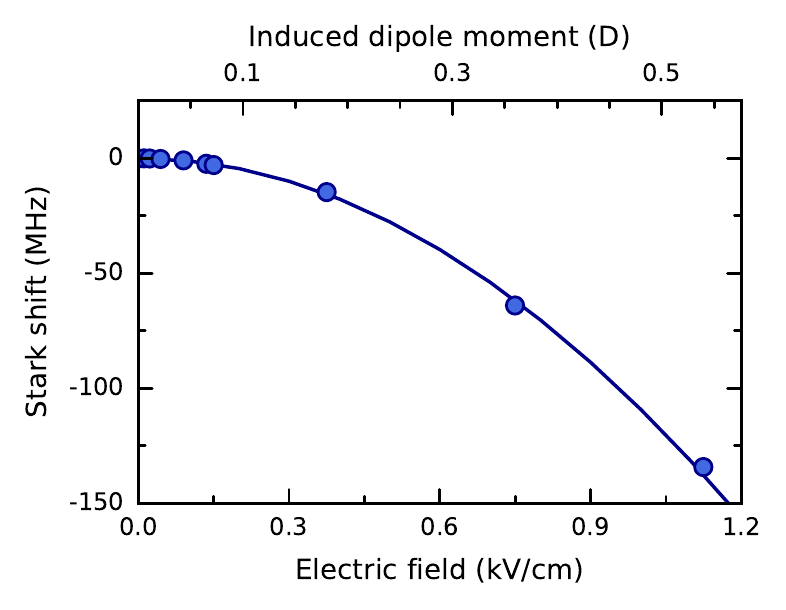} 
\caption{\label{fig:polarize} 
STIRAP at high electric fields. 
Stark shift of the STIRAP transition for various applied electric fields (circles, lower axis).
The applied electric field has been calibrated using a DC Stark shift model and the molecular dipole moment determined in \cite{Gerdes2011}.
The corresponding induced electric dipole moment is given on the upper axis, indicating, that polar molecules with 0.54 D have been produced.
}
\end{figure}

Finally, we polarize the ground state molecules using four rod electrodes within our vacuum system. 
We measure the Stark shift of the ground state transition using STIRAP at different applied voltages. 
With the previously determined dipole moment of 2.72 D for NaK \cite{Gerdes2011}, we calibrate our electric fields and determine the induced dipole moments \cite{Mueller1984}, see Fig. \ref{fig:polarize}.
From the Stark shift we can deduce that polar molecules with dipole moments of up to 0.54 D can be routinely produced by our setup. 
While dark state spectroscopy has already been performed up to dipole moments of 1.06 D \citep{Guo2016}, actual polar molecules have so far only been produced with dipole moments of up to 0.3 D \cite{Park2015}.
We expect to achieve even higher dipole moments with a more stable high voltage power supply.

\section{Conclusion and Outlook}
We have set up a model to describe STIRAP in dipolar molecules beyond a three-level system, taking into account the structure of the Feshbach state as well as the intermediate and the ground state manifolds. 
STIRAP with large one-photon detunings was demonstrated in $^{23}$Na$^{40}$K using the predominantly $v_d=5$ level in the d$^3\Pi$/D$^1\Pi$ coupled system. 
One-photon resonant STIRAP efficiencies of 25 \% and one-photon detuned STIRAP efficiencies of 50 \% were observed for different hyperfine states within the rovibronic ground state.
Our quantitative STIRAP model reproduces the STIRAP efficiencies at different one-photon detunings for all polarization scenarios nicely when laser phase noise is taken into account. 
We find that it is important to include the entire Zeeman (HF) multiplet to accurately describe losses during STIRAP, especially at large one-photon detunings. 
In this case the large pulse area during the transfer can result in significant losses even from far off resonant Zeeman components in the excited state.
The model further encourages the realization of a polarization scenario with $\sigma^+/\sigma^+$-polarized pump/Stokes beams to increase STIRAP efficiency. 
The largest loss in phase-space density occurs in the current setup however already prior to STIRAP due to the low efficiency of Feshbach association.
This could be mitigated in an optical lattice \cite{Covey2016,Reichsollner2016} in the future.
Finally we demonstrated that molecules with a dipole moment of 0.54 D can be created in our setup, the most polar diatomic molecular sample so far. 
We believe that the model developed here can be easily generalized to other molecular species and states, thus allowing for a systematic search for the optimal STIRAP path to the ground state.

\begin{acknowledgments}
We gratefully acknowledge intense discussions with Romain Vexiau and Nadia Bouloufa-Maafa regarding the choice of the intermediate state, support from the DFG (FOR 2247) and the EU (UQUAM).%
\end{acknowledgments}       

\bibliography{references}

\end{document}